\documentclass{article}%
\usepackage{amsmath}
\usepackage{graphicx}
\usepackage{amsfonts}
\usepackage{amssymb}%
\newtheorem{theorem}{Theorem}

\newtheorem{corollary}[theorem]{Corollary}

\newtheorem{definition}[theorem]{Definition}

\newtheorem{proposition}[theorem]{Proposition}
\newtheorem{remark}[theorem]{Remark}

\makeatletter
\@addtoreset{equation}{section}

\makeatother
\begin{document}

\title{Generalized Quantum Turing Machine and its Application to the SAT Chaos Algorithm}
\author{Satoshi Iriyama, Masanori Ohya and Igor Volovich\\Tokyo University of Science\\Depertment of Information Science}
\maketitle

\begin{abstract}
Ohya and Volovich have proposed a new quantum computation model with chaotic
amplification to solve the SAT problem, which went beyond usual quantum
algorithm. In this paper, we generalize quantum Turing machine, and we show in
this general quantum Turing machine (GQTM) that we can treat the Ohya-Volovich
(OV) SAT algorithm.

\end{abstract}

\section{Introduction}

The problem whether NP-complete problems can be P problem has been considered
as one of the most important problems in theory of computational complexity.
Various studies have been done for many years \cite{GPFB}. Ohya and Volovich
\cite{OV1,OV2} proposed a new quantum algorithm with chaotic amplification
process to solve the SAT problem, which went beyond usual quantum algorithm.
This quantum chaos algorithm enabled to solve the SAT problem in a polynomial
time \cite{OV1,OV2,OM}.

In this paper we generalize quantum Turing machine so that it enables to
describe non-unitary evolution of states. This study is based on mathematical
studies of quantum communication channels \cite{O,AO}. It is discussed in this
generalized quantum Turing machine (GQTM) that we can treat the OV SAT algorithm.

In Section 2, we generalize QTM by rewriting usual QTM in terms of channel
transformation so that it contains both dissipative and unitary dynamics. In
Section 3, the SAT problem is reviewed and fundamental quantum unitary gates
are presented. In Section 4, based on the papers \cite{OM,AS}, we concretely
construct the fundamental gates needed for computation of the SAT problem. In
Section 5, we rewrite the total process including a measurement process and
amplifier process with chaotic dynamics by GQTM.

\section{Generalized Quantum Turing Machine}

Classical Turing machine(TM or CTM) $M_{cl}$ is defined by a triplet $\left(
Q,\Sigma,\delta\right)  $, where $\Sigma$ is a finite alphabets with an
identified blank symbol $\#$, $Q$ is a finite set of states (with an initial
state $q_{0}$ and a set of final states $q_{f})$ and $\delta:Q\times
\Sigma\rightarrow Q\times\Sigma\times\left\{  -1,0,1\right\}  $ is a
transition function. Note that $\left\{  -1,0,1\right\}  $ indicates moving
direction of the tape head of TM. The deterministic TM has a deterministic
transition function $\delta:Q\times\Sigma$ $\rightarrow$ $2^{Q}$ $\times
\Sigma$ $\times\left\{  -1,0,1\right\}  ,$ that is, $\delta$ is a
non-branching map, in other words, the range of $\delta$ for each $\left(
q,a\right)  \in Q\times\Sigma$ is unique. A TM $M$ is called non-deterministic
if it is not deterministic.

Quantum Turing machine (QTM) was introduced by Deutsch \cite{D} and has been
extensively studied by Bernstein and Vazirani \cite{BV}.In this section, we
introduce a generalized quantum Turing machine (GQTM), which contains QTM as a
special case.

The Hilbert space $\mathcal{H}$ of QTM consists from complex functions defined
on the space of classical configurations.

\begin{definition}
Usual Quantum Turing machine $M_{q}$ is defined by a quadruplet $M_{q}=\left(
Q,\Sigma,\mathcal{H},U\right)  ,$ where $\mathcal{H}$ is a Hilbert space
described below in (\ref{tur2})and $U$ is a unitary operator on the space
$\mathcal{H}$ of the special form described below in (\ref{tur3}).
\end{definition}

Let $\mathcal{C}=Q\times\Sigma\times\mathbb{Z}$ be the set of all classical
configurations of the Turing machine $M_{cl},$ where $\mathbb{Z}$ is the set
of all integers$.$ It is a countable set and one has%

\begin{equation}
\mathcal{H}=\left\{  \varphi\mid\varphi:\mathcal{C\rightarrow}\mathbb{C}%
;\underset{C\in\mathcal{C}}{\sum}\left\vert \varphi(C)\right\vert ^{2}%
<\infty\right\}  . \label{tur2}%
\end{equation}

Since the configuration $C\in\mathcal{C}$ can be written as $C=$ $\left(
q,A,i\right)  $ one can say that the set of functions $\left\{  \mid
q,A,i>\right\}  $ is a basis in the Hilbert space $\mathcal{H}.$ Here $q\in
Q,$ $i\in\mathbb{Z}$ and $A$ is a function $A:\mathbb{Z\rightarrow\Sigma}.$ We
will call this basis the \textit{computational basis. }

By using the computational basis we now state the conditions to the unitary
operator $U$. We denote the set $\Gamma\equiv\left\{  -1,0,1\right\}  .$ One
requires that there is a function $\delta:Q\times\Sigma\times Q\times
\Sigma\times\Gamma\rightarrow\overset{\sim}{\mathbb{C}}$ which takes values in
the field of computable numbers $\overset{\sim}{\mathbb{C}}$ and such that the
following relation is satisfied:%

\begin{equation}
U\left\vert q,A,i\right\rangle =\underset{p,b,\sigma}{\sum}\delta
(q,A(i),p,b,\sigma)\left\vert p,A_{i}^{b},i+\sigma\right\rangle . \label{tur3}%
\end{equation}
Here the sum runs over the states $p\in Q,$ the symbols $b\in\Sigma$ and the
elements $\sigma\in\Gamma.$ Actually this is a finite sum. The function
$A_{i}^{b}:\mathbb{Z\rightarrow\Sigma}$ is defined as%

\[
A_{i}^{b}(j)=\left\{
\begin{tabular}
[c]{l}%
$b$ if $j=i,$\\
$A(j)$ if $j\neq i.$%
\end{tabular}
\right.
\]

The restriction to the computable number field $\tilde{\mathbb{C}}$ instead of
all the complex number $\mathbb{C}$ is required since otherwise we can not
construct or design a quantum Turing Machine.

Note that if, for some integer $t\in\mathbb{N\equiv}\left\{  1,2,...\right\}
,$ the quantum state $U^{t}\left\vert q_{0},A,0\right\rangle $ is a final
quantum state, i.e. $\left\Vert E_{Q}(q_{F})U^{s}\left\vert q_{0}%
,A,0\right\rangle \right\Vert =1$ and for any $s<t,$ $s\in\mathbb{N}$ one has
$\left\Vert E_{Q}(q_{F})U^{s}\left\vert q_{0},A,0\right\rangle \right\Vert
=0,$ then one says that the quantum Turing machine \textit{halts with running
time }$t$ on input $A.$

Now we define the generalized quantum Turing machine (GQTM) by using of a
channel $\Lambda$ (see below) instead of a unitary operator $U$.

\begin{definition}
Generalized Quantum Turing machine $M_{gq}$ (GQTM) is defined by a quadruplet
$M_{gq}=\left(  Q,\Sigma,\mathcal{H},\Lambda\right)  ,$ where $Q$ and $\Sigma$
are two alphabets, $\mathcal{H}$ is a Hilbert space and $\Lambda$ is a channel
on the space of states on $\mathcal{H}$ of the special form described below.
\end{definition}

Let us explain GQTM in more detailed. GQTM $M_{gq}$ is defined by quadruplet
$\left(  Q,\Sigma,\mathcal{H},\Lambda\right)  $, where $Q$ is a processor
configuration, $\Sigma$ is a set of alphabet including a blank symbol and
$\Lambda$ is a quantum transition function sending a quantum state to a
quantum state. $Q$ and $\Sigma$ are represented by a density operator on
Hilbert space $\mathcal{H}_{Q}$ and $\mathcal{H}_{\Sigma},$which are spanned
by canonical basis $\left\{  \left\vert q\right\rangle ;q\in Q\right\}  $ and
$\left\{  \left\vert a\right\rangle ;a\in\Sigma\right\}  ,$ respectively. A
tape configuration $A$ is a sequence of elements of $\Sigma$ represented by a
density operator on Hilbert space $\mathcal{H}_{\Sigma}$ spanned by a
canonical basis $\left\{  \left\vert A\right\rangle ;A\in\Sigma^{\ast
}\right\}  ,$ where $\Sigma^{\ast}$ is the set of sequences of alphabets in
$\Sigma$. A position of tape head is represented by a density operator on
Hilbert space $\mathcal{H}_{Z}$ spanned by a canonical basis $\left\{
\left\vert i\right\rangle ;i\in\mathbf{Z}\right\}  $. Then a configuration
$\rho$ of GQTM $M_{gq}$ is described by a density operator in $\mathcal{H}%
\equiv\mathcal{H}_{Q}\otimes\mathcal{H}_{\Sigma}\otimes\mathcal{H}_{Z}$. Let
$\mathfrak{S}\left(  \mathcal{H}\right)  $ be the set of all density operators
in Hilbert space $\mathcal{H}$. A quantum transition function $\Lambda$ is
given by a completely positive (CP) channel
\[
\Lambda:\mathfrak{S}\left(  \mathcal{H}\right)  \rightarrow\mathfrak{S}\left(
\mathcal{H}\right)  .
\]
For instance, given a configuration $\rho\equiv\sum_{k}\lambda_{k}\left\vert
\psi_{k}\right\rangle \left\langle \psi_{k}\right\vert ,$ where $\sum
\lambda_{k}=1,\lambda_{k}\geq0$ and $\psi_{k}=\left\vert q_{k}\right\rangle
\otimes\left\vert A_{k}\right\rangle \otimes\left\vert i_{k}\right\rangle \ $
$(q_{k}\in Q,A_{k}\in\Sigma^{\ast},i_{k}\in\mathbf{Z)}$ is a vector in a basis
of $\mathcal{H}.$ This configuration changes to a new configuration
$\rho^{\prime}$ by\ one step transition as $\rho^{\prime}=\Lambda\left(
\rho\right)  =\sum_{k}\mu_{k}\left\vert \psi_{k}\right\rangle \left\langle
\psi_{k}\right\vert $ with $\sum\mu_{k}=1,\mu_{k}\geq0.$

One requirement on GQTM $M_{gq}=\left(  Q,\Sigma,\mathcal{H},\Lambda\right)  $
is the correspondence with QTM. If the channel $\Lambda$ in GQTM will be a
unitary operator $U$\ then GQTM $M_{gq}=\left(  Q,\Sigma,\mathcal{H}%
,\Lambda=U\cdot U^{\ast}\right)  $ reduces to QTM $M_{q}=\left(
Q,\Sigma,\mathcal{H},U\right)  .$

Several studies have been done on QTM whose transition function is represented
by unitary operator, in which various theorems and computational classes in
QTM were discussed in \cite{BV,NO}.

Let us explain how to construct a QTM. Let $\delta$ be a function
\[
\delta:Q\times\Sigma\times Q\times\Sigma\times\left\{  -1,0,1\right\}
\rightarrow\mathbb{C}\mathbf{.}%
\]
For any $q\in Q,a\in\Sigma$, it holds
\[
\sum_{p\in Q,b\in\Sigma,d\in\left\{  -1,0,1\right\}  }\left\vert \delta\left(
q,a,p,b,d\right)  \right\vert ^{2}=1.
\]
For any $q\in Q,a\in\Sigma,$ $q^{\prime}\left(  \neq q\right)  \in
Q,a^{\prime}\left(  \neq a\right)  \in\Sigma$, it holds%

\[
\sum_{p\in Q,b\in\Sigma,d\in\left\{  -1,0,1\right\}  }\delta\left(  q^{\prime
},a^{\prime},p,b,d\right)  ^{\ast}\delta\left(  q,a,p,b,d\right)  =0.
\]

Given QTM $M_{q}$ and its configuration $\rho=\left\vert \varphi\right\rangle
\left\langle \varphi\right\vert $ with $\left\vert \varphi\right\rangle
=\left\vert q,A,i\right\rangle $, after one step, this configuration is
changed by a transition function $\delta$ as%
\begin{align*}
\Lambda_{\delta}(\left\vert q,A,i\right\rangle \langle i,A,q|)  &
=\underset{p,b,\sigma,p^{\prime},b^{\prime},\sigma^{\prime}}{\sum}%
\delta(q,A(i),p,b,\sigma)\delta^{\ast}(q,A(i),p^{^{\prime}},b^{^{\prime}%
},\sigma^{^{\prime}})\\
&  \left\vert p,A_{i}^{b},i+\sigma\right\rangle \langle i+\sigma^{^{\prime}%
},A_{i}^{b^{^{\prime}}},p^{^{\prime}}|
\end{align*}

\begin{remark}
For any $q,p\in Q,a,b\in\Sigma,d\in\left\{  -1,0,1\right\}  $, let
$\delta\left(  q,a,p,b,d\right)  =\left\{  0,1\right\}  $, then QTM is a
reversal TM.
\end{remark}

A transition of GQTM is regarded as a transition of amplitude of each
configuration vector. We categorize \ GQTMs by a property of CP channel
$\Lambda$ as below.

\begin{definition}
A GQTM $M_{gq}$ is called unitary QTM (UQTM, i.e., usual QTM), if all of
quantum transition function $\Lambda$ in $M$ are unitary CP channel.
\end{definition}

For all configuration $\rho=\sum_{n}\lambda_{n}\rho_{n}$ $\left(  \Sigma
_{n}\lambda_{n}=1,\lambda_{n}\geq0\right)  $, a GQTM $M_{gq}$ is called LQTM
$M_{lq}$ if $\Lambda$ is affine ; $\Lambda\left(  \sum_{n}\lambda_{n}\rho
_{n}\right)  =\sum_{n}\lambda_{n}\Lambda\left(  \rho_{n}\right)  .$ Since a
measurement defined by $\Lambda_{M}\rho=%
{\textstyle\sum\limits_{k}}
P_{k}\rho P_{k}$ with a PVM $\left\{  P_{k}\right\}  $ on $\mathcal{H}$ is a
linear CP channel, LQTM may include a measurement process.

For a more general channel the state change is expressed as%

\begin{align*}
\Lambda(\left\vert q,A\left(  i\right)  ,i\right\rangle \langle q,A\left(
i\right)  ,i|)  &  =\underset{p,b,\sigma,p^{\prime},b^{\prime},\sigma^{\prime
}}{\sum}\delta(q,A(i),p,b,\sigma,p^{\prime},b^{\prime},\sigma^{\prime})\\
&  \left\vert p,A_{i}^{b},i+\sigma\right\rangle \langle p,A_{i}^{b^{\prime}%
},i+\sigma|
\end{align*}
with some function $\delta(q,A(i),p,b,\sigma,p^{\prime},b^{\prime}%
,\sigma^{\prime})$ such that the RHS of this relation is a state.

Thus we define two more classes of GQTM for non-unitary CP channels.

\begin{definition}
A GQTM $M_{gq}$ is called a linear QTM(LQTM) if its quantum transition
function $\Lambda$ is a linear quantum channel.
\end{definition}

Unitary operator is linear, hence UQTM is a sub-class of LQTM. moreover,
classical TM is a special class of LQTM.

\begin{definition}
A\ GQTM $M_{gq}$ is called non-linear QTM $\left(  NLQTM\right)  $ if its
quantum transition function $\Lambda$ contains non-linear CP channel.
\end{definition}

A chaos amplifier used in \cite{OV1,OV2} is a non-linear CP channel, the
details of this channel and its application to the SAT problem will be
discussed in the sequel.

\subsection{Computational class for GQTM}

Let us state some language classes which classical Turing machine recognizes.

\begin{definition}
The class of languages is in P if its language is recognized by a
deterministic Turing machine in polynomial time of input size.
\end{definition}

\begin{definition}
The class of languages is in NP if there is a deterministic Turing machine,
called the \textit{verifier, which recognize languages with some informations
in polynomial time} of input size.\textit{ }Besides, if a language $L_{1}$
$\in$NP and $L_{1}$ reduces to $L_{2}\in$NP in polynomial time, a language
$L_{1}$ is NP-complete.
\end{definition}

\begin{definition}
If languages are accepted by non-deterministic Turing machine in polynomial
time of input size with a certain probability, this class of languages are
called the class of bounded probability polynomial time(BPP).
\end{definition}

A NP-complete language is the most difficult one in NP. If there is a
polynomial time algorithm to solve it in the above sense, it implies P=NP. The
existence of such a algorithm is demonstrated in \cite{OV1,OV2} in an extended
quantum domain, as is reviewed in the next section. In this paper we will show
that this OV algorithm can be written by GQTM in the sequel section.

Given a GQTM $M_{gq}=\left(  Q,\Sigma,\delta\right)  $ and an input
configuration $\rho_{0}=\left\vert v_{in}\right\rangle \left\langle
v_{in}\right\vert $, $\left(  \left\vert v_{in}\right\rangle =\left\vert
q_{0}\right\rangle \otimes\left\vert T\right\rangle \otimes\left\vert
0\right\rangle \right)  $, a computation process is described as the following
product of several different types of channels%
\[
\Lambda_{1}\circ\cdots\circ\Lambda_{t}\left(  \rho_{0}\right)  =\rho_{f}%
\equiv\left\vert v_{f}\right\rangle \left\langle v_{f}\right\vert
\]
where $\Lambda_{1},\cdots,\Lambda_{t}$ are CP channels. Applying the CP
channels to an initial state, we obtain a final state $\rho_{f}$ and we
measure this state by a projection (or PVM)%
\[
P_{f}=\left\vert q_{f}\right\rangle \left\langle q_{f}\right\vert \otimes
I_{\Sigma}\otimes I_{Z},
\]
where $I_{\Sigma},I_{Z}$ are identity operators on $\mathcal{H}_{\Sigma
},\mathcal{H}_{Z},$ respectively. Let $p\geq0$ be a halting probability such
that%
\[
tr_{\mathcal{H}_{\Sigma}\otimes\mathcal{H}_{Z}}\left(  P_{f}\rho_{f}\right)
=p\left\vert q_{f}\right\rangle \left\langle q_{f}\right\vert .
\]

Then, we define the \textit{acceptance} (\textit{rejection)} of GQTM and some
classes of languages.

\begin{definition}
Given GQTM $M_{gq}$ and a language $L$, if there exists $N$ steps when we
obtain the configuration of {\normalsize acceptance} $\left(  or\text{
}{\normalsize rejection}\right)  $by the probability $p$, we say that the GQTM
$M_{gq}$ {\normalsize accepts }$\left(  or\text{ }{\normalsize rejects}%
\right)  $$L$ by the probability $p$, and its computational complexity is $t$.
\end{definition}

\begin{definition}
A language $L$ is bounded quantum probability polynomial time GQTM(BGQPP) if
there is a polynomial time GQTM $M_{gq}$ which accepts $L$ with probability
$p\geq\frac{1}{2}$.
\end{definition}

Similarly, we can define the class of languages BUQPP$\left(  =\text{BQPP}%
\right)  $, BLQPP, BNLQPP(=BGQPP) corresponding to UQTM, LQTM and NLQTM, respectively.

In Section 2, it is pointed out that LQTM includes classical TM, which it may
imply: BPP$\subseteq$BLQPPL$\subseteq$BNLQPP$\subseteq$BGQPP. Moreover, if
NLQTM accepts the SAT OV algorithm in polynomial time with probability
$p\geq\frac{1}{2}$, then we may have the inclusion%

\[
NP\subseteq BGQPP
\]
We will discuss this inclusion in Sec. 4 by constructing GQTM which accepts
the SAT OV algorithm.

\section{SAT Problem}

Let $X\equiv\left\{  x_{1},\ldots,x_{n}\right\}  ,n\in\mathbf{N}$ be a set.
$x_{k}$ and its negation $\overline{x}_{k}\left(  k=1,\ldots,n\right)  $ are
called literals Let $\overline{X}\equiv\left\{  \overline{x_{1}}%
,\ldots,\overline{x_{n}}\right\}  $ be a set, then the set of all literals is
denoted by $X^{\prime}\equiv X\cup\overline{X}=\left\{  x_{1},\ldots
,x_{n},\overline{x_{1}},\ldots,\overline{x_{n}}\right\}  $. The set of all
subsets of $X^{\prime}$ is denoted by $\mathcal{F}\left(  X^{\prime}\right)  $
and an element $C\in\mathcal{F}\left(  X^{\prime}\right)  $ is called a
clause. We take a truth assignment to all variables $x_{k}$. If we can assign
the truth value to at least one element of $C$, then $C$ is called
satisfiable. When $C$ is satisfiable, the truth value $t\left(  C\right)  $ of
$C$ is regarded as true, otherwise, that of $C$ is false. Take the truth
values as ''true $\leftrightarrow$1, false $\leftrightarrow$0''. Then $C$is
satisfiable iff $t\left(  C\right)  =1$.

Let $L=\left\{  {0,1}\right\}  $ be a Boolean lattice with usual join $\vee$
and meet $\wedge$, and $t\left(  x\right)  $ be the truth value of a literal
$x$ in $X$. Then the truth value of a clause $C$ is written as $t\left(
C\right)  \equiv\vee_{x\in C}t\left(  x\right)  $.

Moreover the set $\mathcal{C}$ of all clauses $C_{j}\left(  {j=1,2,\cdots
,m}\right)  $ is called satisfiable iff the meet of all truth values of
$C_{j}$ is 1; $t\left(  \mathcal{C}\right)  \equiv\wedge_{j=1}^{m}t\left(
{C_{j}}\right)  =1$. Thus the SAT problem is written as follows:

\begin{definition}
SAT Problem: Given a Boolean set $X\equiv\left\{  {x_{1},\cdots,x_{n}%
}\right\}  $and a set $\mathcal{C}=\left\{  C_{1},\cdots,C_{m}\right\}  $ of
clauses, determine whether $\mathcal{C}$ is satisfiable or not.
\end{definition}

That is, this problem is to ask whether there exists a truth assignment to
make $\mathcal{C}$ satisfiable. It is known in usual algorithm that it is
polynomial time to check the satisfiability only when a specific truth
assignment is given, but we can not determine the satisfiability in polynomial
time when an assignment is not specified.

In \cite{OM} we discussed the quantum algorithm of the SAT problem, which was
rewritten in \cite{AS} with showing that the OM SAT-algorithm is
combinatorial. In \cite{OV1,OV2} it is shown that the chaotic quantum
algorithm can solve the SAT problem in polynomial time.

Ohya and Masuda pointed out \cite{OM} that the SAT problem, hence all other NP
problems, can be solved in polynomial time by quantum computer if the
superposition of two orthogonal vectors $\left|  0\right\rangle $ and $\left|
1\right\rangle $ is physically detected. However this detection is considered
not to be possible in the present technology. The problem to be overcome is
how to distinguish the pure vector $\left|  0\right\rangle $ from the
superposed one $\alpha\left|  0\right\rangle +\beta\left|  1\right\rangle ,$
obtained by the OM SAT-quantum algorithm, if $\beta$ is not zero but very
small. If such a distinction is possible, then we can solve the NPC problem in
the polynomial time. In \cite{OV1,OV2} it is shown that it can be possible by
combining nonlinear chaos amplifier with the quantum algorithm, which implies
the existence of a mathematical algorithm solving NP=P. The algorithm of Ohya
and Volovich is not known to be in the framework of quantum Turing algorithm
or not. This aspect is studied in this paper.

\subsection{Quantum computation}

In this subsection, we review fundamentals of quantum computation (see, for
instance, \cite{OV3}). Let $\mathbb{C}$ be the set of all complex numbers, and
$\left\vert 0\right\rangle $ and $\left\vert 1\right\rangle $ be the two unit
vectors $\binom{1}{0}$ and $\binom{0}{1}$, respectively. Then, for any two
complex numbers $\alpha$ and $\beta$ satisfying $\left\vert \alpha\right\vert
^{2}+\left\vert \beta\right\vert ^{2}=1$, $\alpha\left\vert 0\right\rangle
+\beta\left\vert 1\right\rangle $ is called a qubit. For any positive integer
$N$, let $\mathcal{H}$ be the tensor product Hilbert space defined as $\left(
\mathbb{C}^{2}\right)  ^{\otimes N}$ and let $\left\{  \left\vert
e_{i}\right\rangle ;0\leq i\leq2^{N-1}\right\}  $ be the basis whose elements
are denoted as%

\begin{align*}
\left\vert e_{0}\right\rangle  &  =\left\vert 0\right\rangle \otimes\left\vert
0\right\rangle \cdots\otimes\left\vert 0\right\rangle \equiv\left\vert
0,0,\cdots,0\right\rangle ,\\
\left\vert e_{1}\right\rangle  &  =\left\vert 1\right\rangle \otimes\left\vert
0\right\rangle \cdots\otimes\left\vert 0\right\rangle \equiv\left\vert
1,0,\cdots,0\right\rangle ,\\
\left\vert e_{2}\right\rangle  &  =\left\vert 0\right\rangle \otimes\left\vert
1\right\rangle \cdots\otimes\left\vert 0\right\rangle \equiv\left\vert
0,1,\cdots,0\right\rangle ,\\
&  \vdots\\
\left\vert e_{2^{N}-1}\right\rangle  &  =\left\vert 1\right\rangle
\otimes\left\vert 1\right\rangle \cdots\otimes\left\vert 1\right\rangle
\equiv\left\vert 1,1,\cdots,1\right\rangle .
\end{align*}
For any two qubits $\left\vert x\right\rangle $ and $\left\vert y\right\rangle
$, $\left\vert x,y\right\rangle $ and $\left\vert x^{N}\right\rangle $ is
defined as $\left\vert x\right\rangle \otimes\left\vert y\right\rangle $ and
$\underset{N\text{ times}}{\underbrace{\left\vert x\right\rangle \otimes
\cdots\otimes\left\vert x\right\rangle }}$, respectively.

The usual (unitary) quantum computation can be formulated mathematically as
the multiplication by unitary operators. Let $U_{NOT}$,$U_{CN}$ and $U_{CCN}$
be the three unitary operators defined as%
\begin{align*}
U_{NOT}  &  \equiv\left\vert 1\right\rangle \left\langle 0\right\vert
+\left\vert 0\right\rangle \left\langle 1\right\vert ,\\
U_{CN}  &  \equiv\left\vert 0\right\rangle \left\langle 0\right\vert \otimes
I+\left\vert 1\right\rangle \left\langle 1\right\vert \otimes U_{NOT},\\
U_{CCN}  &  \equiv\left\vert 0\right\rangle \left\langle 0\right\vert \otimes
I\otimes I+\left\vert 1\right\rangle \left\langle 1\right\vert \otimes
\left\vert 0\right\rangle \left\langle 0\right\vert \otimes I+\left\vert
1\right\rangle \left\langle 1\right\vert \otimes\left\vert 1\right\rangle
\left\langle 1\right\vert \otimes U_{NOT}.
\end{align*}
$U_{NOT}$,$U_{CN}$ and $U_{CCN}$ represent the NOT-gate, the Controlled-NOT
gate and the Controlled-Controlled-NOT gate, respectively. Moreover, Hadamard
transformation $H$ is defined as the transformation on $\mathbb{C}^{2}$ such
as%
\[
H\left\vert 0\right\rangle =\frac{1}{\sqrt{2}}\left(  \left\vert
0\right\rangle +\left\vert 1\right\rangle \right)  ,\text{ }H\left\vert
1\right\rangle =\frac{1}{\sqrt{2}}\left(  \left\vert 0\right\rangle
-\left\vert 1\right\rangle \right)  .
\]
These four operators $U_{NOT}$, $U_{CN}$, $U_{CCN}$ and $H$ are called the
elementary gates here. For any $k\in\mathbb{N}$, $U_{H}^{\left(  N\right)
}\left(  k\right)  $ denotes the $k$-tuple Hadamard transformation on $\left(
\mathbb{C}^{2}\right)  ^{\otimes N}$ defined as%

\[
U_{H}^{\left(  N\right)  }\left(  k\right)  \left\vert 0^{N}\right\rangle
=\frac{1}{2^{k/2}}\left(  \left\vert 0\right\rangle +\left\vert 1\right\rangle
\right)  ^{\otimes k}\left\vert 0^{N-k}\right\rangle =\frac{1}{2^{k/2}}%
\sum\limits_{i=0}^{2^{k-1}}\left\vert e_{i}\right\rangle \otimes\left\vert
0^{N-k}\right\rangle .
\]

The above unitary operators can be extended to the unitary operators on
$\left(  \mathbb{C}^{2}\right)  ^{\otimes N}$:%

\begin{align*}
U_{NOT}^{\left(  N\right)  }(u)  &  \equiv I^{\otimes u-1}\otimes\left(
\left\vert 0\right\rangle \left\langle 1\right\vert +\left\vert 1\right\rangle
\left\langle 0\right\vert \right)  I^{\otimes N-u-1}\\
U_{CN}^{\left(  N\right)  }\left(  u,v\right)   &  \equiv I^{\otimes
u-1}\otimes\left\vert 0\right\rangle \left\langle 0\right\vert \otimes
I^{\otimes N-u-1}+I^{\otimes u-1}\otimes\left\vert 1\right\rangle \left\langle
1\right\vert \\
&  \otimes I^{\otimes v-u-1}\otimes U_{NOT}\otimes I^{\otimes N-v-1}\\
U_{CCN}^{\left(  N\right)  }\left(  u,v,w\right)   &  =I^{\otimes u-1}%
\otimes\left\vert 0\right\rangle \left\langle 0\right\vert \otimes I^{\otimes
N-u-1}+I^{\otimes u-1}\otimes\left\vert 1\right\rangle \left\langle
1\right\vert \\
&  \otimes I^{\otimes v-u-1}\otimes\left\vert 0\right\rangle \left\langle
0\right\vert \otimes I^{\otimes N-v-1}\\
&  +I^{\otimes u-1}\otimes\left\vert 1\right\rangle \left\langle 1\right\vert
\otimes I^{\otimes v-u-1}\otimes\left\vert 1\right\rangle \left\langle
1\right\vert \otimes\\
&  I^{\otimes w-t-1}\otimes U_{NOT}\otimes I^{\otimes N-w-1},
\end{align*}
where $u,v$ and $w$ be a positive integers satisfying $1\leq u<v<w\leq N$.

Furthermore we have the following three unitary operators $U_{AND},U_{OR}$ and
$U_{COPY}$ , called the logical gates; (see \cite{AS})%

\begin{align*}
U_{AND}  &  \equiv\sum_{\varepsilon_{1},\varepsilon_{2}\in\left\{
0,1\right\}  }\left\{  \left\vert \varepsilon_{1},\varepsilon_{2}%
,\varepsilon_{1}\wedge\varepsilon_{2}\right\rangle \left\langle \varepsilon
_{1},\varepsilon_{2},0\right\vert +\left\vert \varepsilon_{1},\varepsilon
_{2},1-\varepsilon_{1}\wedge\varepsilon_{2}\right\rangle \left\langle
\varepsilon_{1},\varepsilon_{2},1\right\vert \right\} \\
&  =\left\vert 0,0,0\right\rangle \left\langle 0,0,0\right\vert +\left\vert
0,0,1\right\rangle \left\langle 0,0,1\right\vert +\left\vert
1,0,0\right\rangle \left\langle 1,0,0\right\vert +\left\vert
1,0,1\right\rangle \left\langle 1,0,1\right\vert \\
&  +\left\vert 0,1,0\right\rangle \left\langle 0,1,0\right\vert +\left\vert
0,1,1\right\rangle \left\langle 0,1,1\right\vert +\left\vert
1,1,1\right\rangle \left\langle 1,1,0\right\vert +\left\vert
1,1,0\right\rangle \left\langle 1,1,1\right\vert .
\end{align*}

\begin{align*}
U_{OR}  &  \equiv\sum_{\varepsilon_{1},\varepsilon_{2}\in\left\{  0,1\right\}
}\left\{  \left\vert \varepsilon_{1},\varepsilon_{2},\varepsilon_{1}%
\vee\varepsilon_{2}\right\rangle \left\langle \varepsilon_{1},\varepsilon
_{2},0\right\vert +\left\vert \varepsilon_{1},\varepsilon_{2},1-\varepsilon
_{1}\vee\varepsilon_{2}\right\rangle \left\langle \varepsilon_{1}%
,\varepsilon_{2},1\right\vert \right\} \\
&  =\left\vert 0,0,0\right\rangle \left\langle 0,0,0\right\vert +\left\vert
0,0,1\right\rangle \left\langle 0,0,1\right\vert +\left\vert
1,0,1\right\rangle \left\langle 1,0,0\right\vert +\left\vert
1,0,0\right\rangle \left\langle 1,0,1\right\vert \\
&  +\left\vert 0,1,1\right\rangle \left\langle 0,1,0\right\vert +\left\vert
0,1,0\right\rangle \left\langle 0,1,1\right\vert +\left\vert
1,1,1\right\rangle \left\langle 1,1,0\right\vert +\left\vert
1,1,0\right\rangle \left\langle 1,1,1\right\vert .
\end{align*}

\begin{align*}
U_{COPY}  &  \equiv\sum_{\varepsilon_{1}\in\left\{  0,1\right\}  }\left\{
\left\vert \varepsilon_{1},\varepsilon_{1}\right\rangle \left\langle
\varepsilon_{1},0\right\vert +\left\vert \varepsilon_{1},1-\varepsilon
_{1}\right\rangle \left\langle \varepsilon_{1},1\right\vert \right\} \\
&  =\left\vert 0,0\right\rangle \left\langle 0,0\right\vert +\left\vert
0,1\right\rangle \left\langle 0,1\right\vert +\left\vert 1,1\right\rangle
\left\langle 1,0\right\vert +\left\vert 1,0\right\rangle \left\langle
1,1\right\vert .
\end{align*}
We call $U_{AND},U_{OR}$ and $U_{COPY}$ , AND gate, OR gate and COPY gate,
respectively, whose extensions on $\left(  \mathbb{C}^{2}\right)  ^{\otimes
N}$ are denoted by $U_{AND}^{\left(  N\right)  },U_{OR}^{\left(  N\right)  }$
and $U_{COPY}^{\left(  N\right)  }$, which are expressed as%

\begin{align*}
U_{AND}^{\left(  N\right)  }(u,v,w)  &  =\sum_{\varepsilon_{1},\varepsilon
_{2}\in\left\{  0,1\right\}  }I^{\otimes u-1}\otimes\left\vert \varepsilon
_{1}\right\rangle \left\langle \varepsilon_{1}\right\vert I^{\otimes
v-u-1}\otimes\left\vert \varepsilon_{2}\right\rangle \left\langle
\varepsilon_{2}\right\vert \\
&  I^{\otimes w-v-u-1}\otimes\left\vert \varepsilon_{1}\wedge\varepsilon
_{2}\right\rangle \left\langle 0\right\vert I^{\otimes N-w-v-u}+\\
&  I^{\otimes u-1}\otimes\left\vert \varepsilon_{1}\right\rangle \left\langle
\varepsilon_{1}\right\vert I^{\otimes v-u-1}\otimes\\
&  \left\vert \varepsilon_{2}\right\rangle \left\langle \varepsilon
_{2}\right\vert I^{\otimes w-v-u-1}\otimes\left\vert 1-\varepsilon_{1}%
\wedge\varepsilon_{2}\right\rangle \left\langle 1\right\vert I^{\otimes
N-w-v-u}.
\end{align*}%
\begin{align*}
U_{OR}^{\left(  N\right)  }\left(  u,v,w\right)   &  \equiv\sum_{\varepsilon
_{1},\varepsilon_{2}\in\left\{  0,1\right\}  }I^{\otimes u-1}\otimes\left\vert
\varepsilon_{1}\right\rangle \left\langle \varepsilon_{1}\right\vert
I^{\otimes v-u-1}\otimes\left\vert \varepsilon_{2}\right\rangle \left\langle
\varepsilon_{2}\right\vert \\
&  I^{\otimes w-v-u-1}\otimes\left\vert \varepsilon_{1}\vee\varepsilon
_{2}\right\rangle \left\langle 0\right\vert I^{\otimes N-w-v-u}+\\
&  I^{\otimes u-1}\otimes\left\vert \varepsilon_{1}\right\rangle \left\langle
\varepsilon_{1}\right\vert I^{\otimes v-u-1}\otimes\left\vert \varepsilon
_{2}\right\rangle \left\langle \varepsilon_{2}\right\vert \\
&  I^{\otimes w-v-u-1}\otimes\left\vert 1-\varepsilon_{1}\vee\varepsilon
_{2}\right\rangle \left\langle 1\right\vert I^{\otimes N-w-v-u}.
\end{align*}%
\begin{align*}
U_{COPY}^{\left(  N\right)  }\left(  u,v\right)   &  \equiv\sum_{\varepsilon
_{1}\in\left\{  0,1\right\}  }I^{\otimes u-1}\left\vert \varepsilon
_{1}\right\rangle \left\langle \varepsilon_{1}\right\vert I^{\otimes
v-u-1}\left\vert \varepsilon_{1}\right\rangle \left\langle 0\right\vert
I^{\otimes N-v-u}\\
&  +I^{\otimes u-1}\left\vert \varepsilon_{1}\right\rangle \left\langle
\varepsilon_{1}\right\vert I^{\otimes v-u-1}\left\vert 1-\varepsilon
_{1}\right\rangle \left\langle 1\right\vert I^{\otimes N-v-u}.
\end{align*}
where $u,v$ and $w$ are positive integers satisfying $1\leq u<v<w\leq N$.
These operators can be written, in terms of elementary gates, as%
\begin{align*}
U_{OR}^{\left(  N\right)  }\left(  u,v,w\right)   &  =U_{CN}^{\left(
N\right)  }\left(  u,w\right)  \cdot U_{CN}^{\left(  N\right)  }\left(
v,w\right)  \cdot U_{CCN}^{\left(  N\right)  }\left(  u,v,w\right)  ,\\
U_{AND}^{\left(  N\right)  }\left(  u,v,w\right)   &  =U_{CCN}^{\left(
N\right)  }\left(  u,v,w\right)  ,\\
U_{COPY}^{\left(  N\right)  }\left(  u,v\right)   &  =U_{CN}^{\left(
N\right)  }\left(  u,v\right)  .
\end{align*}

\section{SAT Algorithm}

In this section, we explain the algorithm of the SAT problem which has been
introduced by Ohya-Masuda \cite{OM} and developed by Accardi-Sabbadini
\cite{AS}. The computation of the truth value can be done by by a combination
of the unitary operators on a Hilbert space $\mathcal{H}$, so that the
computation is described by the unitary quantum algorithm. The detail of this
section is given in the papers \cite{OM,OV2,AS,AI}, so we will discuss just
the essence of the OM algorithm. Throughout this section, let $n$ be the total
number of Boolean variables used in the SAT problem. Let $\mathcal{C}$ be a
set of clauses whose cardinality is equal to $m$. Let $\mathcal{H}=\left(
\mathbf{C}^{2}\right)  ^{\otimes n+\mu+1}$ be a Hilbert space and $\left\vert
v_{0}\right\rangle $ be the initial state $\left\vert v_{0}\right\rangle
=\left\vert 0^{n},0^{\mu},0\right\rangle $, where $\mu$ is the number of dust
qubits which is determined by the following proposition. Let $U_{\mathcal{C}%
}^{\left(  n\right)  }$ be a unitary operator for the computation of the SAT:%

\[
U_{\mathcal{C}}^{\left(  n\right)  }\left\vert v_{0}\right\rangle =\frac
{1}{\sqrt{2^{n}}}\sum_{i=0}^{2^{n}-1}\left\vert e_{i},x^{\mu},t_{e_{i}}\left(
\mathcal{C}\right)  \right\rangle \equiv\left\vert v_{f}\right\rangle
\]
where $x^{\mu}$ denotes the $\mu$ strings in the dust bits and $t_{e_{i}%
}\left(  \mathcal{C}\right)  $ is the truth value of $\mathcal{C}$ with
$e_{i}$. In \cite{OM,AS}, $U_{\mathcal{C}}^{\left(  n\right)  }$ was constructed.

Let $\left\{  s_{k};k=1,\dots,m\right\}  $ be the sequence defined as%
\begin{align*}
s_{1}  &  =n+1,\\
s_{2}  &  =s_{1}+card\left(  C_{1}\right)  +\delta_{1,card\left(
C_{1}\right)  }-1,\\
s_{i}  &  =s_{i-1}+card\left(  C_{i-1}\right)  +\delta_{1,card\left(
C_{i-1}\right)  },\text{ \ \ }3\leq i\leq m,
\end{align*}
where $card\left(  C_{i}\right)  $ means the cardinality of a clause $C_{i}$.
And let define $s_{f}$ as%
\[
s_{f}=s_{m}-1+card\left(  C_{m}\right)  +\delta_{1,card\left(  C_{m}\right)
}.
\]
Note that the number $m$ of the clause is at most $2n$. Then we have the
following proposition and theorem \cite{AI}.

\begin{proposition}
For $m\geq2$, the total number of dust qubits $\mu$ is
\begin{align*}
\mu &  =s_{f}-1-n\\
&  =\sum_{k=1}^{m}card\left(  C_{k}\right)  +\delta_{1,card\left(
C_{k}\right)  }-2.
\end{align*}

\end{proposition}

Determining $\mu$ and the work spaces for computing $t\left(  C_{k}\right)  $,
we can construct $U_{\mathcal{C}}^{\left(  n\right)  }$ concretely. We use the
following unitary gates for this concrete expression:%

\[
U_{AND}^{\left(  x\right)  }\left(  k\right)  =\left\{
\begin{array}
[c]{c}%
U_{AND}^{\left(  x\right)  }\left(  s_{k+1}-1,s_{k+2}-2,s_{k+2}-1\right)
,\text{ \ \ \ }1\leq k\leq m-2\\
U_{AND}^{\left(  x\right)  }\left(  s_{m}-1,s_{f}-1,s_{f}\right)  ,\text{
\ \ \ \ }k=m-1
\end{array}
\right.  ,
\]

\begin{align*}
U_{OR}^{\left(  x\right)  }\left(  k\right)   &  =\bar{U}_{OR}^{\left(
x\right)  }\left(  l_{4},s_{k}-card\left(  C_{k}\right)  -1,s_{k}-card\left(
C_{k}\right)  -2\right)  \cdot\cdots\cdot\bar{U}_{OR}^{\left(  x\right)
}\left(  l_{3},s_{k},s_{k}+1\right)  \bar{U}_{OR}^{\left(  x\right)  }\left(
l_{1},l_{2},s_{k}\right)  ,\\
\bar{U}_{OR}^{\left(  x\right)  }\left(  u,v,w\right)   &  =\left\{
\begin{array}
[c]{c}%
U_{OR}^{\left(  x\right)  }\left(  u,v,w\right)  ,\text{ \ \ \ \ \ \ \ \ }%
x_{u}\in C_{k}\\
U_{NOT}^{\left(  x\right)  }\left(  u\right)  \cdot U_{OR}^{\left(  x\right)
}\left(  u,v,w\right)  \cdot U_{NOT}^{\left(  x\right)  }\left(  u\right)
,\text{ \ \ \ \ \ }\bar{x}_{u}\in C_{k}\\
U_{NOT}^{\left(  x\right)  }\left(  u\right)  \cdot U_{NOT}^{\left(  x\right)
}\left(  v\right)  \cdot U_{OR}^{\left(  x\right)  }\left(  u,v,w\right)
\cdot U_{NOT}^{\left(  x\right)  }\left(  u\right)  \cdot U_{NOT}^{\left(
x\right)  }\left(  v\right)  ,\text{ \ \ \ \ \ }\bar{x}_{u},\bar{x}_{v}\in
C_{k}%
\end{array}
\right.  ,
\end{align*}
where $l_{1},l_{2},l_{3},l_{4}$ are positive integers such that $x_{z}\in
C_{k}$ or $\bar{x}_{z}\in C_{k}$, $\left(  z=l_{1},\ldots,l_{4}\right)  $.

\begin{theorem}
The unitary operator $U_{\mathcal{C}}^{\left(  n\right)  }$, is represented as%
\begin{align*}
U_{\mathcal{C}}^{\left(  n\right)  }  &  =U_{AND}^{\left(  n+\mu+1\right)
}\left(  m-1\right)  \cdot U_{AND}^{\left(  n+\mu+1\right)  }\left(
m-2\right)  \cdot\cdots\cdot U_{AND}^{\left(  n+\mu+1\right)  }\left(
1\right) \\
&  \cdot U_{OR}^{\left(  n+\mu+1\right)  }\left(  m\right)  \cdot
U_{OR}^{\left(  n+\mu+1\right)  }\left(  m-1\right)  \cdot\cdots\cdot
U_{OR}^{\left(  n+\mu+1\right)  }\left(  1\right)  \cdot U_{H}^{\left(
n+\mu+1\right)  }\left(  n\right)  .
\end{align*}

\end{theorem}

\subsection{The resulting state in the SAT algorithm}

Applying the above unitary operator to the initial state, we obtain the final
state $\rho.$The result of the computation is registered as $\left\vert
t\left(  \mathcal{C}\right)  \right\rangle $ in the last section of the final
vector, which will be taken out by a projection $P_{n+\mu,1}\equiv I^{\otimes
n+\mu}\otimes\left\vert 1\right\rangle \left\langle 1\right\vert $ onto the
subspace of $\mathcal{H}$ spanned by the vectors $\left\vert \varepsilon
^{n},\varepsilon^{\mu},1\right\rangle $..

The following theorem is easily seen.

\begin{theorem}
$\mathcal{C}$ is SAT if and only if%
\[
P_{n+\mu,1}U_{\mathcal{C}}^{\left(  n\right)  }\left\vert v_{0}\right\rangle
\neq0
\]

\end{theorem}

According to the standard theory of quantum measurement, after a measurement
of the event $P_{n+\mu,1}$, the state $\rho=|v_{f}><v_{f}|$ becomes
\[
\rho\rightarrow\frac{P_{n+\mu,1}\rho P_{n+\mu,1}}{Tr\rho P_{n+\mu,1}%
}=:\overline{\rho}%
\]
Thus the solvability of the SAT problem is reduced to check that $\rho
^{\prime}\neq0$. The difficulty is that the probability
\[
Tr\overline{\rho}P_{n+\mu,1}=\Vert P_{n+\mu,1}\left\vert v_{f}\right\rangle
\Vert^{2}={\frac{|T(\mathcal{C}_{0})|}{2^{n}}}%
\]
is very small in some cases, where $|T(\mathcal{C}_{0})|$ is the cardinality
of the set $T(\mathcal{C}_{0})$, of all the truth functions $t$ such that
$t(\mathcal{C}_{0})=1.$

\emph{We put }$q\equiv$\emph{\ }$\sqrt{{\frac{r}{2^{n}}}}$ \emph{with}
$r\equiv|T(\mathcal{C}_{0})|$ \emph{. Then if }$r$\emph{\ is suitably large to
detect it, then the SAT problem is solved in polynomial time. However, for
small }$r,$\emph{\ the probability is very small so that we in fact do not get
an information about the existence of the solution of the equation }%
$t(C_{0})=1,$\emph{\ hence in such a case we need further deliberation.}

Let go back to the SAT algorithm. After the quantum computation, the quantum
computer will be in the state
\[
\left\vert v_{f}\right\rangle =\sqrt{1-q^{2}}\left\vert \varphi_{0}%
\right\rangle \otimes\left\vert 0\right\rangle +q\left\vert \varphi
_{1}\right\rangle \otimes\left\vert 1\right\rangle
\]
where $\left\vert \varphi_{1}\right\rangle $ and $\left\vert \varphi
_{0}\right\rangle $ are normalized $n$ (=$n+\mu)$ qubit states and
$q=\sqrt{r/2^{n}}.$ Effectively our problem is reduced to the following $1$
qubit problem: The above state $\left\vert v_{f}\right\rangle $ is reduced to
the state
\[
\left\vert \psi\right\rangle =\sqrt{1-q^{2}}\left\vert 0\right\rangle
+q\left\vert 1\right\rangle ,
\]
and we want to distinguish between the cases $q=0$ and $q>0$(small positive
number). Let us denote the correspondence from $\rho_{0}$ $\equiv\left\vert
v_{0}\right\rangle \left\langle v_{0}\right\vert $ with $\rho$ by a channel
$\Lambda_{I};$ $\rho=\Lambda_{I}\rho_{0}.$

\quad It is argued in \cite{BBBV} that quantum computer can speed up
\textbf{NP} problems quadratically but not exponentially. The no-go theorem
states that if the inner product of two quantum states is close to 1, then the
probability that a measurement distinguishes which one of the two is
exponentially small. And one may claim that amplification of this
distinguishability is not possible in usual quantum algorithm. At this point
we emphasized \cite{OV2} that we do not propose to make a measurement which
will be overwhelmingly likely to fail. What we did is a proposal to use the
output $\left\vert \psi\right\rangle $ of the quantum computer as an input for
another device which uses chaotic dynamics. \emph{The amplification would be
not possible if we use the standard model of quantum computations with a
unitary evolution. However the idea of the paper \cite{OV1,OV2} is different.
In \cite{OV1,OV2} it is proposed to combine quantum computer with a chaotic
dynamics amplifier. Such a quantum chaos computer is a new model of
computations and we demonstrate that the amplification is possible in the
polynomial time.}

\quad One could object that we do not suggest a practical realization of the
new model of computations. But at the moment nobody knows of how to make a
practically useful implementation of the standard model of quantum computing
ever. It seems to us that the quantum chaos computer considered in \cite{OV2}
deserves an investigation and has a potential to be realizable.

\subsection{Chaotic dynamics\qquad}

Various aspects of classical and quantum chaos have been the subject of
numerous studies (\cite{O,OV3} and ref's therein). Here we will briefly review
how chaos can play a constructive role in computation (see \cite{OV1,OV2} for
the details).

Chaotic behavior in a classical system usually is considered as an exponential
sensitivity to initial conditions. It is this sensitivity we would like to use
to distinguish between the cases $q=0$ and $q>0$ discussed in the previous subsection.

Consider the so called logistic map which is given by the equation
\[
x_{n+1}=ax_{n}(1-x_{n})\equiv g(x),~~~x_{n}\in\left[  0,1\right]  .
\]

\noindent\noindent\noindent The properties of the map depend on the parameter
$a.$ If we take, for example, $a=3.71,$ then the Lyapunov exponent is
positive, the trajectory is very sensitive to the initial value and one has
the chaotic behavior \cite{OV2}. It is important to notice that if the initial
value $x_{0}=0,$ then $x_{n}=0$ for all $n.$

\quad The state $\left\vert \psi\right\rangle $ of the previous subsection is
transformed into the density matrix of the form
\[
\overline{\rho}=q^{2}P_{1}+\left(  1-q^{2}\right)  P_{0}%
\]
where $P_{1}$ and $P_{0}$ are projectors to the state vectors $\left\vert
1\right\rangle $ and $\left\vert 0\right\rangle .$ One has to notice that
$P_{1}$ and $P_{0}$ generate an Abelian algebra which can be considered as a
classical system. The density matrix $\rho$ above is interpreted as the
initial data, and we apply the channel $\Lambda\equiv\Lambda_{CA}$ due to the
logistic map as%
\[
\Lambda_{CA}\left(  \overline{\rho}\right)  =\frac{\left(  I+g\left(
\overline{\rho}\right)  \sigma_{3}\right)  }{2},
\]
where $I$ is the identity matrix and $\sigma_{3}$ is the z-component of Pauli
matrices.
\[
\overline{\rho}_{k}=\Lambda_{CA}^{k}\left(  \overline{\rho}\right)
\]
\ To find a proper value $k$ we finally measure the value of $\sigma_{3}$ in
the state $\rho_{k}$ such that%

\[
M_{k}\equiv tr\overline{\rho}_{k}\sigma_{3}.
\]
We obtain \cite{OV2}

\begin{theorem}
\quad%
\[
\overline{\rho}_{k}=\frac{(I+g^{k}(q^{2})\sigma_{3})}{2},\text{ and }%
M_{k}=g^{k}(q^{2}).
\]

\end{theorem}

\quad Thus the question is whether we can find such a $k$ in polynomial steps
of $n\ $satisfying the inequality $M_{k}\geq\frac{1}{2}$ for very small but
non-zero $q^{2}.$ Here we have to remark that if one has $q=0$ then
$\overline{\rho}=P_{0}$ and we obtain $M_{k}=0$ for all $k.$ If $q\neq0,$ the
chaotic dynamics leads to the amplification of the small magnitude $q$ in such
a way that it can be detected. The transition from $\overline{\rho}$ to
$\overline{\rho}_{k}$ is nonlinear and can be considered as a classical
evolution because our algebra generated by $P_{0}$ and $P_{1}$ is abelian. The
amplification can be done within at most 2n steps due to the following
propositions. Since $g^{k}(q^{2})$ is $x_{k}$ of the logistic map
$x_{k+1}=g(x_{k})$ with $x_{0}=q^{2},$ we use the notation $x_{k}$ in the
logistic map for simplicity.

\begin{theorem}
For the logistic map $x_{n+1}=ax_{n}\left(  1-x_{n}\right)  $ with
$a\in\left[  0,4\right]  $ and $x_{0}\in\left[  0,1\right]  $, let $x_{0}$ be
$\frac{1}{2^{n}}$ and a set $J$ be $\left\{  0,1,2,\dots,n,\dots,2n\right\}
$. If $a$ is $3.71$, then there exists an integer $k$ in $J$ satisfying
$x_{k}>\frac{1}{2}.$
\end{theorem}

\begin{theorem}
Let $a$ and $n$ be the same in above theorem. If there exists $k$ in $J$ such
that $x_{k}>\frac{1}{2},$ then $k>\frac{n-1}{\log_{2}3.71-1}.$
\end{theorem}

\begin{corollary}
\label{Cor1}If $x_{0}\equiv\frac{r}{2^{n}}$ with $r\equiv\left\vert T\left(
\mathcal{C}\right)  \right\vert $ and there exists $k$ in $J$ such that
$x_{k}>\frac{1}{2},$ then there exists $k$ satisfying the following inequality
if $\mathcal{C}$ is SAT.
\[
\left[  \frac{n-1-\log_{2}r}{\log_{2}3.71-1}\right]  \leq k\leq\left[
\frac{5}{4}\left(  n-1\right)  \right]  .
\]

\end{corollary}

From these theorems, for all $k$, it holds%
\[
M_{k}\left\{
\begin{array}
[c]{c}%
=0\text{ \ \ \ iff }\mathcal{C}\text{ is not SAT}\\
>0\text{ \ \ \ iff }\mathcal{C}\text{ is SAT \ \ \ \ }%
\end{array}
\right.
\]

\section{SAT algorithm in GQTM}

In this section, we construct a GQTM for the OV SAT algorithm. The GQTM with
the chaos amplifier belongs to NLQTM because the chaos amplifier is
represented by non-linear CP channel. The OV algorithm runs from an initial
state $\rho_{0}\equiv\left\vert v_{0}\right\rangle \left\langle v_{0}%
\right\vert $ to $\overline{\rho}_{k}$ through $\rho\equiv\left\vert
v_{f}\right\rangle \left\langle v_{f}\right\vert .$ The computation from
$\rho_{0}\equiv\left\vert v_{0}\right\rangle \left\langle v_{0}\right\vert $
to $\rho\equiv\left\vert v_{f}\right\rangle \left\langle v_{f}\right\vert $ is
due to unitary channel $\Lambda_{C}\equiv U_{C}\bullet U_{C},$ and that from
$\rho\equiv\left\vert v_{f}\right\rangle \left\langle v_{f}\right\vert $ to
$\overline{\rho}_{f}$ is due to a non-unitary channel $\Lambda_{CA}^{k}%
\circ\Lambda_{I},$ so that all computation can be done by $\Lambda_{CA}%
^{k}\circ\Lambda_{I}\circ\Lambda_{C},$ which is a completely positive, so the
whole computation process is deterministic. It is a multi-track (actually 4
tracks) GQTM that represents this whole computation process.

A multi-track GQTM has some workspaces for calculation, whose tracks are
independent each other. This independence means that the TM can operate only
one track at one step and all tracks do not affect each other. Let us explain
our computation by a multi-track GQTM. The first track stores the input data
and the second track stores the value of literals. The third track is used for
the computation of $t\left(  C_{i}\right)  ,\left(  i=1,\cdots,m\right)  $
described by unitary operators. The fourth track is used for the computation
of $t\left(  \mathcal{C}\right)  $ denoting the result. The work of GQTM is
represented by the following 8 steps:

\begin{itemize}
\item Step 1 : Store the counter $c=0$ in Track 1. Calculate $\left[  \frac
{5}{4}\left(  n-1\right)  \right]  +1$, we take this value as the maximum
value of the counter. Then, store it in Track 4.

\item Step 2 : Calculate $c+1$ and store it in Track 4.

\item Step 3 : Apply the Hadamard transform to Track 2.

\item Step 4 : Calculate $t\left(  C_{1}\right)  ,\cdots t\left(
C_{m}\right)  $ and store them in Track 3.

\item Step 5 : Calculate $t\left(  \mathcal{C}\right)  $ by using the value of
the third track, and store $t\left(  \mathcal{C}\right)  $ in Track 4.

\item Step 6 : Empty the first, second and third Tracks.

\item Step 7 : Apply the chaos amplifier to the result state obtained up to
the step 6.

\item Step 8 : If $c=\left[  \frac{5}{4}\left(  n-1\right)  \right]  +1$ or
GQTM is in the final state, GQTM halts. If GQTM is not in the final state,
GQTM runs the step 2 to the step 8 again.
\end{itemize}

Let us explain the above steps for unitary computation (OM algorithm; i.e., up
to the steps 6 above) by an example. Let the number of literals be $n$ and
that of clauses be $m$. Then the language is represented by the following strings%

\[
0^{n}X%
{\textstyle\prod_{i=1}^{m}}
C_{S}G\left(  C_{i}\right)  C_{E},
\]
where%

\[
G\left(  C_{i}\right)  =\varepsilon_{1}\varepsilon_{2}\ldots\varepsilon
_{n}Y\overline{\varepsilon_{1}}\overline{\varepsilon_{2}}\ldots\overline
{\varepsilon_{n}}%
\]

\[
\varepsilon_{k}=\left\{
\begin{array}
[c]{c}%
0\text{ \ \ }k\notin I_{i}\\
1\text{ \ \ }k\in I_{i}%
\end{array}
\right.
\]

\[
\overline{\varepsilon_{k}}=\left\{
\begin{array}
[c]{c}%
0\text{ \ \ }k\notin I_{i}^{\prime}\\
1\text{ \ \ }k\in I_{i}^{\prime}%
\end{array}
\right.
\]
and $X,C_{S},Y,$ $C_{E}$ are used as particular symbols of clauses. For
example, given $X=\left\{  1,2,3\right\}  ,$ $\mathcal{C}=\left\{  C_{1}%
,C_{2},C_{3}\right\}  ,$ $C_{1}=\left(  \left\{  1,2\right\}  ,\left\{
3\right\}  \right)  ,$ $C_{2}=\left(  \left\{  3\right\}  ,\left\{  2\right\}
\right)  ,C_{3}=\left(  \left\{  1\right\}  ,\left\{  2,3\right\}  \right)  ,$
the input tape will be%

\[
000XC_{S}110Y001C_{E}C_{S}001Y010C_{E}C_{S}100Y011C_{E}%
\]

First, our GQTM applies DFT to a part of literals on the track 2. The
transition function for DFT is written by the following table. Put the vector
in $\mathcal{H}_{Q}$ by $q_{\cdot}$ instead of $\left\vert q_{\cdot
}\right\rangle $ and denote the direction moving the tape head by $R$ for the
right and $L$ to the left (Note that $O$ is the starting position).

\begin{center}
$%
\begin{array}
[c]{lllll}
& \# & 0 & 1 & X\\
q_{0} &  & q_{a},0,R & q_{a},1,R & \\
q_{a} &  & q_{a},0,R & q_{a},1,R & q_{b},X,L\\
q_{b} & q_{f},\#,R & \frac{1}{\sqrt{2}}q_{b},0,L+\frac{1}{\sqrt{2}}\varphi
_{b},1,L & \frac{1}{\sqrt{2}}q_{b},0,L-\frac{1}{\sqrt{2}}q_{b},1,L &
\end{array}
$
\end{center}

The tape head moves to the right until it reads a symbol $C_{S}$. When the
tape head reads $C_{S}$, GQTM increases a program counter by one, while moves
to the right until it reads $1$. Then GQTM stops increasing the counter and
the tape head moves to the top of the tape. According to the program counter,
the tape head moves to the right as reducing the counter by one. When the
counter becomes zero, GQTM reads the data and calculates OR with the data in
the track 2, then GQTM writes the result in the track 3. GQTM goes back to the
top of the track 1 and repeats the above processes until it reads $Y$.

When GQTM reads $Y$, it calculates OR with the negation and repeats the
processes as above. When it reads $C_{E}$, it writes down $f_{C_{k}}$ in the
track 3 and clean the workspace for the next calculation. Then GQTM reads the
blank symbol $\#$, and it begins to calculate AND. The calculation of AND is
done on the track 4. GQTM calculates them as moving to the left because the
position of the tape head is at the end of the track 3 when the OR calculation
is finished. Then the result of the calculation is showed on the top of the
track 4.

The transition function of OR calculation is described, similar as classical
TM, by the following three tables:

\begin{center}%
\begin{tabular}
[c]{lllllll}
& $0$ & $1$ & $C_{S}$ & $X$ & $Y$ & $\#$\\
$q_{0}$ & $q_{a},0,R$ &  & $q_{b},C_{S},R$ &  & $q_{\overline{OR}},Y,R$ & \\
$q_{a}$ & $q_{a},0,R$ & $q_{a},1,R$ & $q_{b},C_{S},R$ & $q_{a},X,R$ &
$q_{\overline{OR}},Y,R$ & $q_{AND},\#,N$\\
$q_{b}$ & $q_{b,1},0,R$ & $q_{c,1},0,L$ &  &  &  & \\
$q_{b,1}$ & $q_{b,2},0,R$ & $q_{c,1},0,L$ &  &  &  & \\
$\vdots$ &  &  &  &  &  & \\
$q_{b,k}$ & $q_{b,k+1},0,R$ & $q_{c,k+1},0,L$ &  &  &  & \\
$\vdots$ &  &  &  &  &  & \\
$q_{b,n-1}$ & $q_{b,n},0,R$ & $q_{c,n},0,L$ &  &  &  & \\
$q_{b,n}$ &  &  &  &  & $q_{\overline{OR}},Y,R$ & \\
$q_{c,1}$ & $q_{c,1},0,L$ & $q_{c,1},1,L$ & $q_{c,1},C_{S},L$ & $q_{c,1},X,L$
&  & $q_{d,1},\#,R$\\
$\vdots$ &  &  &  &  &  & \\
$q_{c,n}$ & $q_{c,n},0,L$ & $q_{c,n},1,L$ & $q_{c,n},C_{S},L$ & $q_{c,n},X,L$
&  & $q_{d,n},\#,R$\\
$q_{d,1}$ & $q_{t2,0},0,N$ & $q_{t2,1},1,N$ &  &  &  & \\
$q_{d,2}$ & $q_{d,1},0,R$ & $q_{d,1},1,R$ &  &  &  & \\
$\vdots$ &  &  &  &  &  & \\
$q_{d,k}$ & $q_{d,k-1},0,R$ & $q_{d,k-1},1,R$ &  &  &  & \\
$\vdots$ &  &  &  &  &  & \\
$q_{d,n}$ & $q_{d,n-1},0,R$ & $q_{d,n-1},0,R$ &  &  &  &
\end{tabular}

\begin{tabular}
[c]{llllllll}
& $0$ & $1$ & $Y$ & $C_{S}$ & $C_{E}$ & $X$ & $\#$\\
$q_{\overline{OR}}$ & $q_{g,1,},0,R$ & $q_{h,1},0,L$ &  &  &  &  & \\
$q_{e}$ & $q_{e},0,R$ &  & $q_{g},Y,R$ & $q_{e},C_{S},R$ &  &  & \\
$q_{g}$ & $q_{g,1},0,R$ & $q_{h,1},0,L$ &  &  &  &  & \\
$q_{g,1}$ & $q_{g,2},0,R$ & $q_{h,2},0,L$ &  &  &  &  & \\
$\vdots$ &  &  &  &  &  &  & \\
$q_{g,k}$ & $q_{g,k+1},0,R$ & $q_{h,k+1},0,L$ &  &  &  &  & \\
$\vdots$ &  &  &  &  &  &  & \\
$q_{g,n-1}$ & $q_{g,n},0,R$ & $q_{h,n},0,L$ &  &  &  &  & \\
$q_{g,n}$ &  &  &  &  & $q_{j},0,L$ &  & \\
$q_{h,1}$ & $q_{h,1},0,L$ & $q_{h,1},1,L$ & $q_{h,1},Y,L$ & $q_{h,1},C_{S},L$
&  & $q_{h,1},X,L$ & $q_{i,1},\#,R$\\
$\vdots$ &  &  &  &  &  &  & \\
$q_{h,n}$ & $q_{h,n},0,L$ & $q_{h,n},1,L$ & $q_{h,n},Y,L$ & $q_{h,n},C_{S},L$
&  & $q_{h,n},X,L$ & $q_{i,n},\#,R$\\
$q_{i,1}$ & $q_{t2,1},0,N$ & $q_{t2,0},1,N$ &  &  &  &  & \\
$q_{i,2}$ & $q_{i,1},0,R$ & $q_{i,1},1,R$ &  &  &  &  & \\
$\vdots$ &  &  &  &  &  &  & \\
$q_{i,k}$ & $q_{i,k-1},0,R$ & $q_{i,k-1},1,R$ &  &  &  &  & \\
$\vdots$ &  &  &  &  &  &  & \\
$q_{i,n}$ & $q_{i,n-1},0,R$ & $q_{i,n-1},0,R$ &  &  &  &  & \\
$q_{j}$ & $q_{j},0,L$ &  & $q_{j},0,L$ & $q_{t2,a},0,N$ &  &  &
\end{tabular}

\begin{tabular}
[c]{llll}
& $0$ & $1$ & $\#$\\
$q_{t3,0}$ & $q_{t3,0},0,R$ & $q_{t3,1},1,R$ & $q_{a},0,N$\\
$q_{t3,1}$ & $q_{t3,1},0,R$ & $q_{t3,1},1,R$ & $q_{a},1,N$\\
$q_{t3,a}$ & $q_{t4,0},0,N$ & $q_{t4,1},1,N$ & \\
$q_{t3,b}$ & $q_{t3,b},\#,L$ & $q_{t3,b},\#,L$ & $q_{t3,c},\#,R$\\
$q_{t3,c}$ &  &  & $q_{a},0,N$%
\end{tabular}

\end{center}

The transition function of AND calculation is described by the following table:

\begin{center}%
\begin{tabular}
[c]{llll}
& $0$ & $1$ & $\#$\\
$q_{t4,0}$ &  &  & $q_{t3,b},0,R$\\
$q_{t4,1}$ &  &  & $q_{t3,b},1,R$\\
$q_{AND}$ &  &  & $q_{t4,a},\#,L$\\
$q_{t4,a}$ & $q_{t4,a},\#,L$ & $q_{t4,b},\#,L$ & $q_{t4,c},\#,R$\\
$q_{t4,b}$ & $q_{t4,a},\#,L$ & $q_{t4,b},\#,L$ & $q_{t4,d},\#,R$\\
$q_{t4,c}$ &  &  & $q_{f},0,L$\\
$q_{t4,d}$ &  &  & $q_{f},1,R$%
\end{tabular}

\end{center}

Let $q_{6}$ be the processor state of GQTM after the step 6 and $T_{i}%
,i=1,\dots,4$ be the strings of the $i$-th track. Then the OM algorithm showed
that the computation of the SAT problem of the example given above gives us
the resulting state $\rho_{6}$ expressed as%
\begin{align*}
\rho_{6}  &  =q^{2}\left\vert q_{6}\right\rangle \left\langle q_{6}\right\vert
\otimes\left\vert T_{1},T_{2},T_{3},T_{4}\left(  1\right)  \right\rangle
\left\langle T_{1},T_{2},T_{3},T_{4}\left(  1\right)  \right\vert
\otimes\left\vert O\right\rangle \left\langle O\right\vert \\
&  +\left(  1-q^{2}\right)  \left\vert q_{6}\right\rangle \left\langle
q_{6}\right\vert \otimes\left\vert T_{1},T_{2},T_{3},T_{4}\left(  0\right)
\right\rangle \left\langle T_{1},T_{2},T_{3},T_{4}\left(  0\right)
\right\vert \otimes\left\vert O\right\rangle \left\langle O\right\vert ,
\end{align*}
where $T_{4}\left(  1\right)  $ (resp. $T_{4}\left(  0\right)  )$ indicates
that the value in the track 4 is $1$ (resp. $0$).

Next step, as the three tracks (1,2,3) can be empty, we can apply the chaos
amplifier to the above $\rho_{6}$ in the following manner:

The transition function of the step 7 denoted by the chaos amplifier is
formally written as
\begin{align*}
\Lambda_{CA}^{\ast k}\left(  \rho_{6}\right)   &  =g^{k}\left(  q^{2}\right)
\left\vert q_{7}\right\rangle \left\langle q_{7}\right\vert \otimes\left\vert
T_{4}\left(  1\right)  \right\rangle \left\langle T_{4}\left(  1\right)
\right\vert \otimes\left\vert O\right\rangle \left\langle O\right\vert \\
&  +\left(  1-g^{k}\left(  q^{2}\right)  \right)  \left\vert q_{7}%
\right\rangle \left\langle q_{7}\right\vert \otimes\left\vert T_{4}\left(
0\right)  \right\rangle \left\langle T_{4}\left(  0\right)  \right\vert
\otimes\left\vert O\right\rangle \left\langle O\right\vert
\end{align*}
where $g$ is the logistic map explained in Section 4.2. According to
\ref{Cor1}, GQTM halts in at most $\left[  \frac{5}{4}\left(  n-1\right)
\right]  $ steps with the probability $p\geq\frac{1}{2},$ by which we can
claim that $\mathcal{C}$ is SAT.

\subsection{Computational complexity of the SAT algorithm}

We define the computational complexity of the OV SAT algorithm as the product
of $T_{Q}\left(  U_{\mathcal{C}}^{\left(  n\right)  }\right)  $ and
$T_{CA}\left(  n\right)  ,$where $T_{Q}\left(  U_{\mathcal{C}}^{\left(
n\right)  }\right)  $ is the complexity of unitary computation and
$T_{CA}\left(  n\right)  $ is that of chaos amplification.

The following theorem is essentially discussed in \cite{IA,OV2,OM}.

\begin{theorem}
For a set of clauses $\mathcal{C}$ and $n$ Boolean variables, the
computational complexity of the OV SAT algorithm including the chaos
amplifier, denoted by $T\left(  \mathcal{C},n\right)  $, is obtained as
follows.%
\[
T_{GQTM}\left(  \mathcal{C},n\right)  =T_{Q}\left(  U_{\mathcal{C}}^{\left(
n\right)  }\right)  T_{CA}\left(  n\right)  =\mathcal{O}\left(  poly\left(
n\right)  \right)  ,
\]
where $poly\left(  n\right)  $ denotes a polynomial of $n$.
\end{theorem}

The computational complexity of quantum computer is determined by the total
number of logical quantum gates. This inequality implies that the
computational complexity of SAT algorithm is bounded by $\mathcal{O}\left(
n\right)  $ for the size of input $n$ while a classical algorithm is bounded
by $\mathcal{O}\left(  2^{n}\right)  .$

\subsection{Acknowledgment}

One (MO) of the authors thanks IIAS and SCAT for finatial supports.

\end{document}